\documentclass[12pt]{elsart}
\usepackage{graphicx,epsfig}
\begin{document}
\begin{frontmatter}
\title{How robust is a 2SC quark matter phase under compact star constraints?}
\author[rostock]{D. N. Aguilera}
\author[bielefeld,dubna]{D. Blaschke}
\author[rostock,yerevan]{H. Grigorian
}
\address[rostock]{Institut f\"ur Physik, Universit\"at Rostock, 
D-18051 Rostock, Germany}
\address[bielefeld]{Fakult\"at f\"ur Physik, Universit\"at Bielefeld,
D-33615 Bielefeld, Germany}
\address[dubna]{Bogoliubov  Laboratory of Theoretical Physics,\\
        Joint Institute for Nuclear Research, 141980, Dubna, Russia}
\address[yerevan]{Department of Physics, Yerevan State University,
375025 Yerevan, Armenia}
\begin{abstract}
We study the phase structure and equation of state for two-flavor quark 
matter at low temperature under compact star constraints within a nonlocal 
chiral quark model.
We find that the occurence of a two-flavor color superconducting (2SC) phase
is sensitive to variations of both the formfactor of the interaction and the 
ratio $\eta$ between the coupling constants in the  diquark and the scalar 
meson channels. 
Our study suggests that for standard values of the coupling ratio 
$0.5<\eta<0.75$ either the 2SC phase does not
occur (Gaussian formfactor) or it exists only in a mixed phase with normal 
quark matter (NQ-2SC) 
with a volume fraction less than $20 - 40~\%$, occuring at high
baryon chemical potentials  $\mu_B >1200$ MeV and most likely not relevant
for compact stars. We also present the relevant region of the phase diagram for 
compact star applications and obtain that no gapless 2SC occurs 
at low temperatures. 

\noindent PACS number(s):  04.40.Dg, 12.38.Mh, 26.60.+c, 97.60.Jd
\end{abstract}
\end{frontmatter}
\newpage
\section{Introduction}

One of the challenges of strong interaction physics nowadays is 
to understand the phase diagram of quantum chomodynamics (QCD). 
Particular interest is devoted to the high density 
and low temperature region because of possible applications for the physics of
compact stars.
These investigations became very exciting since it has been reported that 
the energy gaps of color superconducting quark matter could be as big as 
$\Delta\sim 100$ MeV \cite{Rapp:1997zu,Alford:1997zt} and thus being of the 
same order 
as the constituent quark masses dynamically generated in the chiral symmetry
breaking transition. Gaps this large could not only have a direct influence 
on the quark matter equation of state (EoS) but, due to the BCS relation 
$T_c = 0.57 \Delta$ critical temperatures would imply that  
quark matter if formed early in the protoneutron star evolution should be
color superconducting. 
 
Dense and cold matter has been broadly studied for the flavor symmetric case 
(for reviews, see \cite{Rajagopal:2000wf,Alford:2001dt}).
From first principles, the limit of  asymptotic densities can be studied and 
it was found that the color-flavor-locking (CFL) phase, in which  
quarks of all three colors and flavors participate in the condensation,  
is energetically favorable (see \cite{Schafer:1999jg}).
Using the simple Nambu-Jona-Lasinio model \cite{Nambu:1961tp}, it was found 
that for intermediate densities the 2SC phase was favored 
\cite{Buballa:2001gj}. 
Therefore, a typical phase diagram calculated for the flavor symmetric case 
shows the low temperature and high density domain being 
in one of the color superconducting quark matter phases. 

However, for compact star applications the situation is much more complex: 
equilibrium w.r.t. weak interactions ($\beta$-equilibrium), color and charge 
neutrality impose additional constraints to dense matter.
Alford and Rajagopal pointed out the difficulty of the 2SC phase to achieve 
charge neutrality  and therefore they claimed that this phase would not be 
realized in compact stars \cite{Alford:2002kj}. 
Without solving the gap equation, they treated the strange quark mass as a 
parameter in their calculations and concluded  that the CFL 
phase would be favorable against 2SC. 

More recently, Steiner, Reddy and Prakash \cite{Steiner:2002gx} found 
within an NJL 
model calculation that the 2SC phase may be realized only in a very small 
region of the phase diagram at lower densities which eventually would be 
covered by the hadronic phase if it was properly take into account. 
Again, the 2SC phase seemed not to be reliable,  basically because the 
charge neutrality condition imposes a strong constraint to the quark chemical 
potentials. 

The discussion of the existence of a mixed phase of 2SC quark matter and normal quark matter (NQ-2SC) by increasing the asymmetry between the  up and down quarks number densities 
was first performed by Bedaque \cite{Bedaque:1999nu}. 
Within the NJL model, a detailed study of possible neutral mixed phases in 
$\beta$-equilibrium has been performed by Neumann, Buballa and Oertel
\cite{Neumann:2002jm} considering global neutrality between the phases. 
In accordance with the earlier expectations, these authors  showed that at 
very low densities - where the model is 
probably to fail - the normal quark matter component is dominant, at 
intermediate ones the 2SC phase and at still higher densities, the CFL phase. 
Moreover, the latter phase and the strange quarks is supossed to appear at very high densities (nearly above $1500$ MeV \cite{Phase:2005,Buballa:2003qv})  that barely 
reached in the very inner core of compact stars and thus leading to no observable effects.

Unfortunately, in the intermediate density region we have no rigorous 
calculations and we have to rely on phenomenological models suitable, however,
for the investigation of qualitative features of the phase diagram. 
The above mentioned and widely used NJL has the great advantage that chiral 
symmetry breaking and diquark condensation are treated at the same mean-field 
level as  phase transitions of an interacting field theory. 

Besides well-known limitations of the NJL-type models (no confinement,
non-renormalizable) there is some arbitrariness in the choice of the model
parameters even if the strategy is followed that vacuum properties 
of low-lying hadrons should be reproduced (e.g. the pion mass, pion decay 
constant and the chiral condensate), see e.g. 
\cite{Klevansky:1992qe,Buballa:2003qv} for reviews. 


In this work we use a chiral quark model that can represent the nonlocality 
of the quark interactions in a more realistic way via formfactor functions. 
Furthermore, we vary the softness of these functions and 
analyze its 
consequences in the phase structure. 
We investigate the parameter regime and obtain that 
no pure 2SC itself does occur under compact 
star constraints for low temperatures (below $25$ MeV). 
We also study the relevant region of the phase diagram for compact star 
applications and obtain that 
neither gapless 2SC (g2SC \cite{Shovkovy:2003uu}) exists; 
its occurence is limited to a narrow window 
for higher temperatures (above 40 MeV) before the second order phase 
transition to the deconfined phase takes place.

For rather strong  coupling constants in the diquark channel 
($\eta \geq$ 0.86) a mixed phase NQ-2SC is likely to occur in the interior of compact stars. 
For particular values of $\eta$ the corresponding equation of state  has been 
already succesfully used for hybrid star configurations, explaining the 
mass-radius relation of very compact objects like 
RX~J185635-3754 \cite{Grigorian:2003vi} or a huge release of energy in the 
evolution of a protoneutron star with trapped antineutrinos 
\cite{Aguilera:2002dh}. 
 
Nevertheless, for weak and intermediate 
coupling constants (even for $\eta=0.75$, the value suggested by the Fierz 
transformation of a massive gluon exchange or two-flavor instanton induced 
interaction \cite{Buballa:2003qv}), the 2SC phase is unlike to occur. 
If this is the case, this result could have important consequences for the 
neutron star phenomenology: if 2SC does not occur inside compact stars, then 
other patterns of diquark pairing satisfying color and charge neutrality can 
become energetically more favorable than normal quark matter.
Examples are spin-1 condensates, e.g. color-spin-locking (CSL) phases 
\cite{Schafer:2000tw,Schmitt:2003xq,Iwasaki:1994ij} 
The influence of the small gaps ($10 - 100$ keV) 
that are expected for these phases on compact star cooling has recently 
been investigated and they could realize the appropriate cooling phenomenology 
\cite{Grigorian:2004jq}.

\section{Quark matter under compact stars contraints}
\subsection{Two flavor quark matter with color superconductivity}

We use a nonlocal 
chiral model for quark matter in the 2SC phase  
to study dense neutral matter in compact stars  
following \cite{Grigorian:2003vi}. 
Formfactor functions $g(p)$ model the nonlocality of the quark 
interaction in the momentum space. We assume that this 
four fermion interaction is instantaneous and therefore the formfactors 
depend only on the modulus of the three-momentum $p=|\vec{p}|$.  
$G_1$ and $G_2$ are 
the coupling constants in the scalar meson and the diquark channels, 
respectively. While $G_1$ is fixed together with the 
current quark mass and 
range parameters of the formfactors (see below) from hadron observables,
$G_2$ is a free parameter of the approach which we vary using the parameter
$\eta=G_2 / G_1$.  

In the mean field approximation  
the grand canonical thermodynamic potential
is a function of the temperature $T$  and 
of the chemical potentials $\mu_{fc}$ for the quark with flavor 
$f$ ($f\in\{u,d\}$) and color $c$ ($c\in\{r,b,g\}$) given by
\begin{eqnarray}
\Omega_q(\{\mu_{fc}\},T)&=&
-T\sum_n\!\int\! \frac{d^3p}{(2\pi)^3}\;
\frac12 {\rm Tr}\ln\left(\frac{1}{T}\tilde S^{-1}(i\omega_n,\vec{p})\right)
+ V -\Omega_{{\rm vac}}~,
\label{Omega_q}
\end{eqnarray}
where the effective potential
\begin{eqnarray}
V&=& \frac{\phi_u^2+\phi_d^2}{8~G_1} + \frac{\Delta^2}{4~G_2} 
\end{eqnarray}
is a function of 
the order parameters of the theory: 
the mass gap $\phi_f$ and  the
diquark gap $\Delta$.  
The constant $\Omega_{{\rm vac}}$ is chosen such that the
pressure of the physical vacuum vanishes. 
The Matsubara frequencies for fermions are given by $\omega_n=(2n+1)\pi T$
and the Nambu-Gorkov inverse quark propagator can be obtained as 
\begin{equation}
\tilde S^{-1}(p_0,\vec{p})=
\left(
\begin{array}{cc}
\not\!p - \hat M -\hat{\mu}\gamma_0~~~&
\Delta \gamma_5
\varepsilon \epsilon^b
g(p)\\
-\Delta^\ast \gamma_5
\varepsilon \epsilon^b
 g(p)&
\not\!p - \hat M+\hat{\mu}\gamma_0
\end{array}
\right)~,
\end{equation}
where  $\hat{\mu}=\mathrm{diag}(\mu_{fc})$ is the chemical 
potential matrix and the elements of 
$\hat{M}(p)=\mathrm{diag}(M_f(p))$ are 
the dynamical masses of the quarks given by $M_f(p)=m_f+g(p)\phi_f$. 
Here, $\gamma_5$ is the usual 
matrix in the Dirac space and 
$(\varepsilon)^{ik}\equiv \varepsilon^{ik}$ and 
$(\epsilon^b)^{\alpha \beta} \equiv \epsilon^{\alpha \beta b}$ 
are antisymmetric tensors 
in the flavor and color spaces respectively.  

The nonlocality of the interaction between the quarks
in both channels quark-antiquark ($q\bar{q}$) and quark-quark ($qq$) 
is implemented via the same formfactor functions
$g(p)$.
In our calculations we use the Gaussian (G), Lorentzian (L) and cutoff (NJL)
formfactors defined as
\begin{eqnarray}
\label{GF}
g_{\rm G}(p) &=& \exp(-p^2/\Lambda_{\rm G}^2)~,\\
\label{LF}
g_{\rm L}(p) &=& [1 + (p/\Lambda_{\rm L})^2]^{-1},\\
\label{NF}
g_{\rm NJL}(p) &=& \theta(1 - p/\Lambda_{\rm NJL})~.
\end{eqnarray}
The parameter sets (quark mass $m=m_u=m_d$, coupling constant $G_1$, interaction range
$\Lambda$) for the above formfactor models (see Tab. \ref{par}) are
fixed by the pion mass $m_{\pi}=140$  MeV, pion decay constant $f_{\pi}=93$
MeV and the mass gap $\phi=\phi_u=\phi_d=330$ MeV
at $T=\mu=0$ \cite{Schmidt:1994di}.

In order to describe asymmetric two flavor quark matter 
 \cite{Kiriyama:2001ud} in a neutral phase \cite{Huang:2002zd} 
we introduce 
the baryon chemical potential $\mu_B$,  
the chemical potential for the isospin asymmetry 
$\mu_I$ and the chemical potential for the color asymmetry $\mu_8$. 
These chemical potentials are  
 conjugate to the conserved quantities in dense matter in neutron stars: baryon number density, 
electric and color charge density, respectively
(see next subsection).   
We express $\{\mu_{fc}\}$ in terms of $\{\mu_B,\mu_I,\mu_8\}$:
\begin{eqnarray}
\mu_{ur}&=&\mu_{ug}=\frac{1}{3}(\mu_B+4\mu_I+\mu_8)~,\\
\mu_{dr}&=&\mu_{dg}=\frac{1}{3}(\mu_B-2\mu_I+\mu_8)~,\\
\mu_{ub}&=&         \frac{1}{3}(\mu_B+4\mu_I-2\mu_8)~,\\
\mu_{db}&=&         \frac{1}{3}(\mu_B-2\mu_I-2\mu_8)~.
\label{chemicalpotentials}
\end{eqnarray}

For the 2SC pairing pattern we assume that
$\Delta_c=g(p)\Delta(\delta_{c,r}+\delta_{c,g})$ in which the red ($r$) and 
the green ($g$) quarks are paired. 
The dispersion relation of the unpaired color (blue) is  
$E_f(p)=\sqrt{p^2+M^2_f(p)}$. 

As has been shown in \cite{Frank:2003ve} for the 2SC phase the relation
$\phi_u=\phi_d=\phi$ holds and therefore $E_u(p)=E_d(p)=E(p)$ 
so that
the quark thermodynamic potential can be written more explicitly as
\begin{eqnarray} \label{Omeg1}
&&
\Omega_q(\mu_B,\mu_I,\mu_8,T)+\Omega_{vac}
=\frac{\phi^2}{4G_1}+\frac{\Delta^2}{4G_2}
\nonumber\\
&&
-\frac{1}{\pi^2}\int^\infty_0dpp^2\{
\omega\left[\epsilon_b\left(E(p)-\frac{1}{3}(\mu_B+\mu_I)+\frac{2}{3}\mu_8\right)-\mu_I,T\right]+
\nonumber\\
&&
\omega\left[\epsilon_b\left(E(p)+\frac{1}{3}(\mu_B+\mu_I)-\frac{2}{3}\mu_8\right)-\mu_I,T\right]+
\nonumber\\
&&
\omega\left[\epsilon_b\left(E(p)-\frac{1}{3}(\mu_B+\mu_I)+\frac{2}{3}\mu_8\right)+\mu_I,T\right]+
\nonumber\\
&&
\omega\left[\epsilon_b\left(E(p)+\frac{1}{3}(\mu_B+\mu_I)-\frac{2}{3}\mu_8\right)+\mu_I,T\right]
\}
\nonumber\\
&&
-\frac{2}{\pi^2}\int^\infty_0dpp^2\{
\omega\left[\epsilon_r\left(E(p)-\frac{1}{3}(\mu_B+\mu_I)-\frac{1}{3}\mu_8\right)-\mu_I,T\right]+
\nonumber\\
&&
\omega\left[\epsilon_r\left(E(p)+\frac{1}{3}(\mu_B+\mu_I)+\frac{1}{3}\mu_8\right)-\mu_I,T\right]+
\nonumber\\
&&
\omega\left[\epsilon_r\left(E(p)-\frac{1}{3}(\mu_B+\mu_I)-\frac{1}{3}\mu_8\right)+\mu_I,T\right]+
\nonumber\\
&&
\omega\left[\epsilon_r\left(E(p)+\frac{1}{3}(\mu_B+\mu_I)+\frac{1}{3}\mu_8\right)+\mu_I,T\right]
\}~,
\label{ome9}
\end{eqnarray}
where following the notation in \cite{Grigorian:2003vi}
\begin{eqnarray}
&&\epsilon_c(\xi)=\xi\sqrt{1+\Delta^2_c/\xi^2}~,\\
&&
\omega\left[{\rm x},T\right]= T\ln\left
[1+\exp\left(-\frac{{\rm x}}{T}\right)\right]+\frac{{\rm x}}{2}~.
\end{eqnarray}
The factor $2$ in the last integral in (\ref{ome9}) comes from the degeneracy of the
red and green colors ($\epsilon_r(\xi)=\epsilon_g(\xi)$).

The total thermodynamic potential $\Omega$ contains the quark
contribution
$\Omega_q$ and the contribution $\Omega^{id}$ of the leptons 
$l$
that 
are treated as a massless, ideal Fermi gas:
\begin{eqnarray}
\label{omega_tot}
\Omega(\mu_B,\mu_I,\mu_8,\mu_l,T)&=&
\Omega_q(\mu_B,\mu_I,\mu_8,T) 
+\sum_{l} \Omega^{id}(\mu_l,T)~.
\end{eqnarray}

The conditions for the local extremum of $\Omega_q$ correspond to
coupled gap equations for the two order parameters $\phi$ and
$\Delta$, $\partial \Omega / \partial \phi=\partial \Omega /\partial \Delta=0$.
The global minimum of $\Omega_q$ represents the state of
thermodynamic equilibrium from which the equation of state can be
obtained by derivation.

\subsection{Beta equilibrium, charge and color neutrality}

The stellar matter in the quark core of compact stars consists of
$\{u,d\}$ quarks and leptons 
$\{e,\nu_e,\bar \nu_e, \mu,\nu_{\mu},\bar \nu_{\mu}\}$
  under the conditions of

\begin{itemize}
\item
$\beta$-equilibrium:
\begin{eqnarray}
d& \longleftrightarrow & u+e+\bar \nu_e\nonumber\\
d& \longleftrightarrow & u+\mu+\bar \nu_{\mu}\nonumber\\
u+e& \longleftrightarrow & d+ \nu_e~,
\end{eqnarray}
which in terms of chemical potentials reads 
($\mu_{\bar \nu_e}=-\mu_{\nu_e}$, 
$\mu_{\bar \nu_{\mu}}= -\mu_{\nu_{\mu}} \simeq 0$)
\begin{eqnarray}
\mu_e+\mu_{\bar \nu_e} =\mu_{\mu}=-2\mu_I~ ,
\label{B_equil}
\end{eqnarray}
\item
charge neutrality: 
\begin{eqnarray}
\frac{2}{3}n_u-\frac{1}{3}n_d-n_e -n_{\mu}= 0~,
\end{eqnarray}
which could also be written as
\begin{eqnarray}
n_B+n_I-2n_e-n_{\mu}=0~~ 
\end{eqnarray}
and
\item
color neutrality:  
\begin{eqnarray}
n_8=\frac{1}{3}(n_{r}+n_{g}-2n_{b})=0~.
\end{eqnarray}
\end{itemize}
Here the number densities $n_j$ are defined in relation to the corresponding 
chemical potentials $\mu_j$ as
\begin{eqnarray}
&&
n_j=-\frac{\partial \Omega}{\partial \mu_j}\bigg|_{\phi_0,\Delta_0;T}~,
\end{eqnarray}
where the index $j$ denotes the particle species and 
$n_f=\sum_{c}n_{fc}$ and $n_c=\sum_f n_{fc}$.
The solution of the color neutrality condition shows that
$\mu_8$ is about 5-7 MeV in the region of relevant densities
($\mu_B\simeq 900-1500$ MeV). Since the isospin asymmetry is independent
of $\mu_8$ we consider $\mu_{8}\simeq 0$ in our following calculations.

\begin{figure}[htb]
  \begin{center}
    \includegraphics[width=1.3\linewidth,
height=1.0\linewidth,
angle = -90]{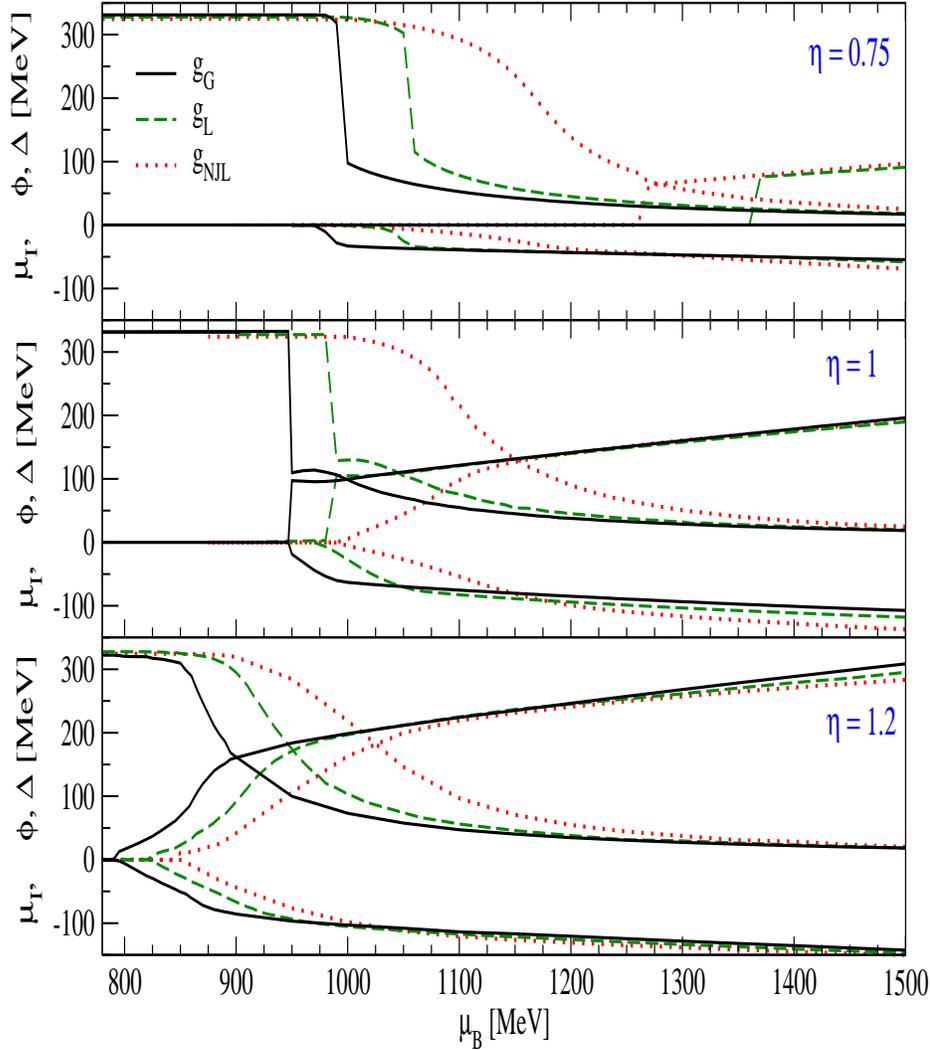}
     \vspace*{-0.5cm}
    \caption{Gap equation solutions for neutral matter in $\beta$-equilibrium 
for different strengths $\eta$ of the coupling constant in the diquark channel
at $T=0$.
In the upper part of each plot, the order parameters $\phi$ and $\Delta$ are 
displayed as a function of the baryon chemical potential. 
In the lower part, the corresponding asymmetry is shown. 
The results are calculated for three different formfactors of the quark 
interaction.}
     \vspace*{0.5cm}
    \label{GEquations}
  \end{center}
\end{figure}

\subsection{Gap equation solutions}

\begin{figure}[htb]
  \begin{center}
    \includegraphics[width=0.7\linewidth,angle = -90]{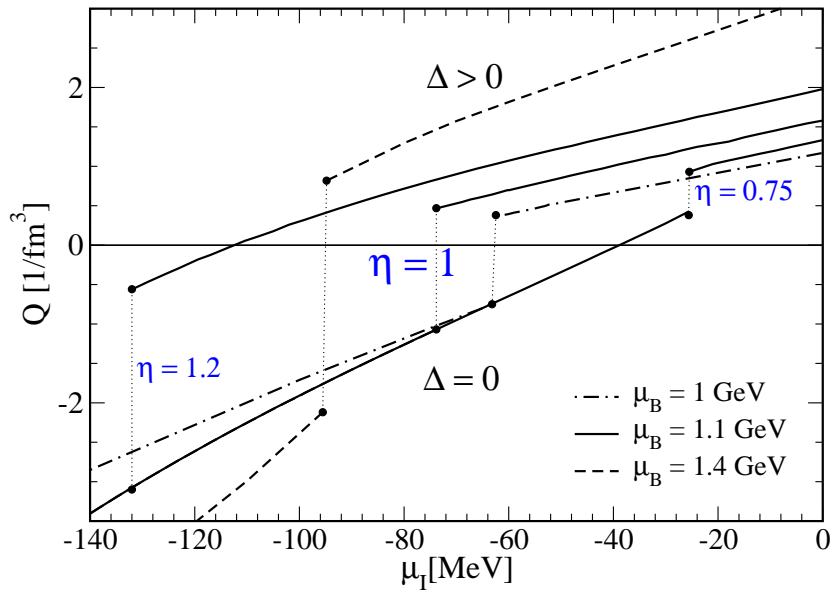}
     \vspace*{-0.5cm}
    \caption{Electric charge density for  2SC ($\Delta >0$) and normal ($\Delta=0$) quark matter phases as a function of isospin chemical potential $\mu_I$ along lines of fixed  baryon chemical potential $\mu_B$ 
(see legend inside figure) for three different coupling constants: $\eta=0.75$ (with the jump on the right), $\eta=1$ (jumps in the middle) and   $\eta=1.2$ (jumps on the left). Gaussian formfactor and $T=0$ are considered. See text for further explanation.}
     \vspace*{0.5cm}
    \label{Q_branches}
  \end{center}
\end{figure}

We solve the gap equations for stellar matter under the constraints of the 
previous subsection. The results are plotted in the Fig. \ref{GEquations}  
for three different strengths $\eta$ of the coupling constant in the diquark 
channel at zero temperature.
For the commonly used value of $\eta = 0.75$ the solutions with Lorentzian 
and NJL formfactors show a normal quark matter phase at intermediate 
densities and a superconducting phase shifted to higher densities.  
The solution with Gaussian formfactor exhibits no diquark condensation at all.
For $\eta = 1$ superconductivity appears immediately at the chiral 
restoration with large gaps for the three formfactors. 
However, in this case a mixed phase construction is necessary to neutralize 
the matter (see next section and Fig. \ref{Q_branches}). 
For an even stronger diquark coupling constant, $\eta = 1.2$, a pure 
superconducting phase with large diquark gaps occurs and  the threshold for the
chiral restoration transition is shifted to lower densities than for  
$\eta=0.75$ and $\eta=1.0$ for the three formfactors. 

\subsection{Mixed phase construction}

To achieve the charge neutrality condition we apply a mixed phase 
construction between the 
subphase with diquark condensation and the subphase of normal quark matter 
 using the  Gibbs conditions for phase equilibrium. The 
temperature $T$, the chemical potentials   
$\{\mu_i\}$ and the pressure $P$ should be equal for the subphases 
involved, in particular, 
\begin{eqnarray}
P_{\Delta=0}(\mu_B,\mu_I,\mu_e,T)&=&P_{\Delta>0}(\mu_B,\mu_I,\mu_e,T)
\end{eqnarray}
where the subscripts denote the normal ($\Delta=0$) and the 
superconducting ($\Delta> 0$) subphases.

In the 2SC mixed phase we construct, the charge neutrality is satisfied as a global 
constraint: the positively charge superconducting phase coexists with the negatively charged normal quark matter phase in an homogeneous mixture. 
Surface tension and Coulomb effects are calculated to be small enough  
not to affect the results qualitatively   
(up to a value of $2 $ MeV/fm$^3$ on the pressure \cite{Reddy:2004my}) and 
we disregard them in this work.

The charges of the subphases (states corresponding to the local minima in 
the thermodynamic potential, i.e. solutions of the gap equations) are shown in
Fig. \ref{Q_branches} for selected values of $\mu_B$ as a function of $\mu_I$ 
for the Gaussian formfactor at zero temperature.  
The upper (lower) branch corresponds 
to the subphase with (without) diquark condensation, the jump represents the 
transition between the subphases where 
the end points are states with the same pressure. 
For a given  $\mu_B$, we can see that increasing the asymmetry $|\mu_I|$ in 
the system (the mismatch between the up and down quark Fermi seas) the diquark 
condensation is disfavored. 
When comparing the charges for different $\eta$, we found that for 
 $\eta=1$ (three lines with jumps that intersect the $Q=0$ line), 
the superconducting branches are positively charged and the 
normal ones are negatively charged. 
At $Q=0$ we obtain the coexistence of the subphases at the same pressure 
and the occurrence of a mixed phase.
For $\eta=0.75$, (at fixed $\mu_B=1.1$ MeV) the lower branch crosses the 
$Q=0$ line and neutral quark matter is just normal.  
At the same $\mu_B$ but in case of the very strong coupling constant  
$\eta=1.2$  we obtain charge neutrality for a pure superconducting phase.  
\begin{figure}[htb]
  \begin{center}
    \includegraphics[width=0.7\linewidth,angle = -90]{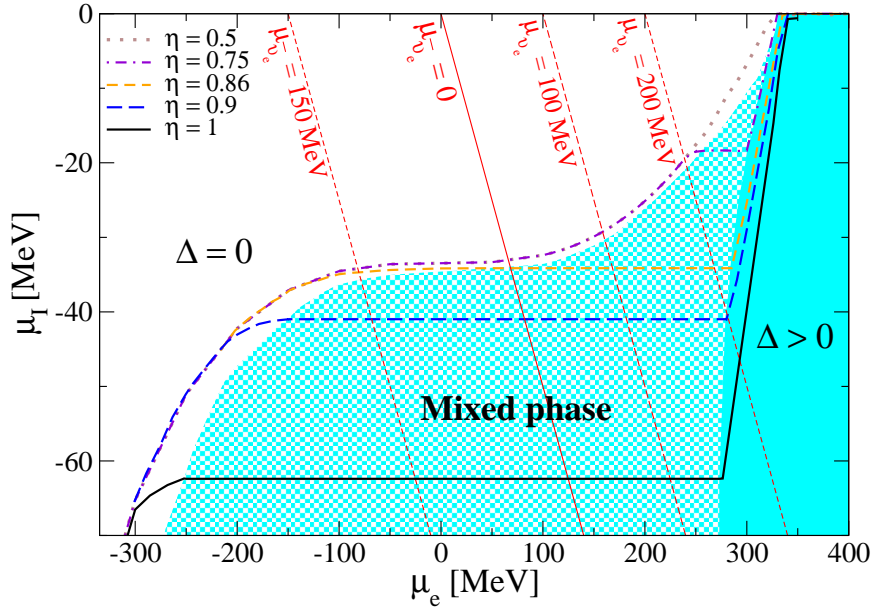}
     \vspace*{-0.5cm}
    \caption{ Solutions of the gap equations and the charge neutrality 
condition (thick lines crossing the figure) in the $\mu_I$ vs. 
$\mu_e$ plane for fixed $\mu_B=1$ GeV and different values of  $\eta$ 
with the Gaussian formfactor at $T=0$. At fixed $\eta$, two branches are shown: 
states with diquark condensation on the right (grey area, $\Delta>0$) 
and states of  normal quark matter (white area, $\Delta=0$) on the left. 
The hatched region in between corresponds to a mixed phase of normal and 
superconducting quark matter. 
The lines corresponding to $\beta$-equilibrium condition are also shown 
(solid and dashed thin vertical lines) for different values of the 
neutrino and antineutrino chemical potential. 
Stellar matter should satisfy both conditions
(intersection of the corresponding lines). 
If $\mu_{\nu_e}=0$,  a mixed phase is preferable for $\eta\geq 0.86$. 
For lower values of the coupling constant diquark condensation is not 
possible in compact stars.}
     \vspace*{0.5cm}
    \label{phase_diag_muI}
  \end{center}
\end{figure}

\begin{figure}[htb]
  \begin{center}
    \includegraphics[width=0.7\linewidth,angle = -90]{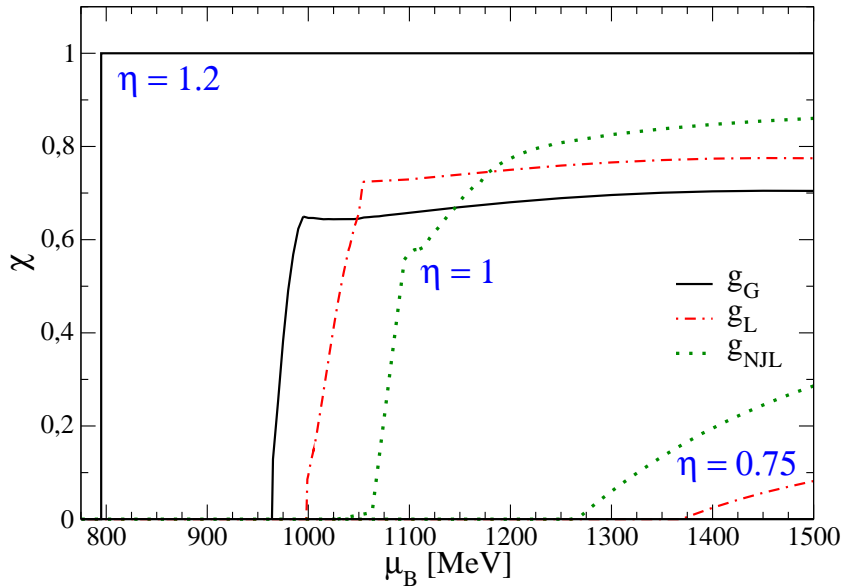}
     \vspace*{-0.5cm}
    \caption{Volume fraction $\chi$ of the 2SC phase as a function of the 
baryon chemical potential obtained by a Glendenning construction for the 
charged-neutral NQ-2SC mixed phase. Results are shown for three different 
diquark coupling constant $\eta$ and three formfactors of the quark 
interaction at $T=0$.}
    \label{chi}
     \vspace*{0.5cm}
  \end{center}
\end{figure}

To explore the relation between asymmetry and $\beta$-equilibrium in neutral 
matter we plot in Fig. \ref{phase_diag_muI} the solution of the gap equations 
under the condition $Q=0$ in the $\mu_I$ vs. $\mu_e$ plane
for the Gaussian formfactor at $T=0$. 
The results are shown for different values of $\eta$ and the lines for 
the $\beta$-equilibrium (\ref{B_equil}) correspond to different values 
of the chemical potential for the  neutrinos and antineutrinos.  
For a  fixed $\eta$, e.g. $\eta=1$, 
two branches are shown: states in a  phase with diquark condensation 
on the right 
($\Delta>0$, for large $\mu_e$) and states in a normal quark matter phase 
($\Delta=0$, for large negative $\mu_e$) on the left. 
The plateau in between them shows that the system  undergoes   
a phase transition from the superconducting to the normal phase 
(and the two subphases coexist in a mixed phase) when the asymmetry increases 
or the number of electrons decreases.
The stellar matter in compact objects 
should satisfy both conditions which are simultaneously
fulfilled at the intersection of the corresponding lines.
Therefore, for the situation in which there are no trapped leptons in the 
star ($\mu_{\nu_e}=0$), a mixed phase is preferable for $\eta\geq 0.86$ 
for the Gaussian formfactor. 
For lower values of the coupling constant diquark condensation is not possible.
A similar analysis shows, for the two other formfactors, that the values of 
$\eta$ at which superconductivity is possible are smaller but its appearance 
is shifted to higher densities, see Fig. \ref{chi}.
\begin{figure}[hbt]
  \begin{center}
    \includegraphics[width=0.7\linewidth,angle = -90]{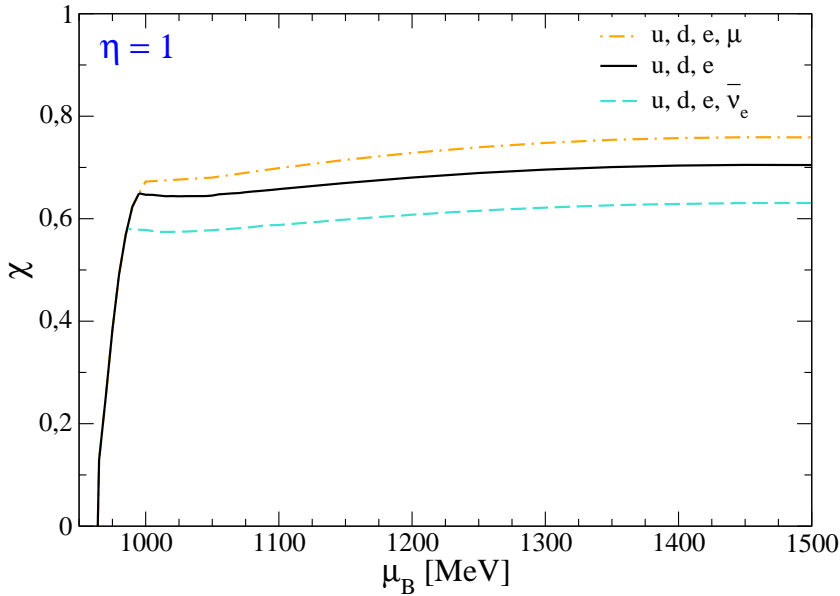}
     \vspace*{-0.5cm}
    \caption{Same as figure \ref{chi} but here
 the Gaussian formfactor is used to analyze the contribution of the 
leptons participating in the $\beta$-equilibrium for $\eta=1$ .  
The solid line corresponds to a system of $u,d$ quarks and electrons $e$. 
The dash-dotted line shows the increase of $\chi$ if also muons $\mu$ 
are included and the dashed line shows the reduction of $\chi$ if 
trapped antineutrinos are considered with $\mu_{\bar \nu_e}=200$ MeV. 
}
    \label{chi_lepton}
     \vspace*{0.5cm}
  \end{center}
\end{figure}

The volume fraction that is occupied by the subphase with diquark condensation 
is defined as
\begin{eqnarray}
\chi = Q_{\Delta>0}/(Q_{\Delta>0}-Q_{\Delta=0})
\label{chi_eq}
\end{eqnarray}
and is plotted in Fig. \ref{chi} as a function of  $\mu_B$ at $T=0$. 
For a strong coupling like $\eta=1.2$, we obtain that $\chi=1$ once quark matter appears 
and therefore a pure superconducting phase for the three formfactors is 
realized. 
For intermediate strength of $\eta=1$, the fraction of the superconducting 
subphase is between 0.6 and 0.9 depending on the formfactor and on the 
density. 
For  $\eta=0.75$, we obtain that no diquark condensation is possible for the 
Gaussian formfactor and for the Lorentzian and NJL formfactors the percentage 
of superconducting matter is less than $50\%$, whereby the onset of color
superconductivity is shifted to higher baryon chemical potentials and thus
possibly becomes irrelevant for compact stars.

In Fig. \ref{chi_lepton} the contribution of the leptons to the volume 
fraction is shown. While muons are negatively charged and act in favor of the 
charge neutrality - increasing the fraction of diquark condensation -  
trapped antineutrinos increment the asymmetry in the system and produce the 
opposite effect. 

\subsection{Phase diagram}

\begin{figure}[htb]
  \begin{center}
    \includegraphics[width=0.7\linewidth,angle = -90]{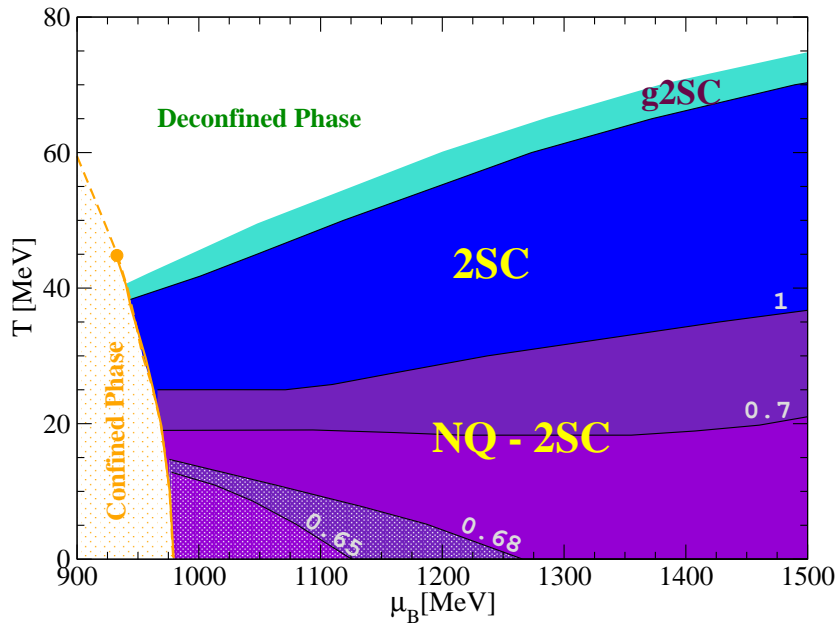}
     \vspace*{-0.5cm}
    \caption{
Relevant region of the phase diagram  for compact stars applications 
for two flavor neutral quark matter within a nonlocal chiral model 
with Gaussian formfactor. 
For  strong coupling constant $\eta=1$, 
a rich structure of color superconducting quark matter phases
is found above the first order chiral transition (solid line on the left).  
For the quark matter sector we obtain  
the occurence of a mixed phase NQ-2SC, a pure 2SC, a g2SC and a second order 
phase transition to a deconfined phase 
as the temperature increases.   
Solid lines within the NQ-2SC region caracterize 
equal values of the volume fraction $\chi$ of the 2SC phase 
indicated by numbers over the corresponding lines.
For $\eta=0.75$ this structure of superconducting phases 
 disappears and we obtain normal quark matter above the 
chiral phase transition.}
    \label{phasediag}
     \vspace*{0.5cm}
  \end{center}
\end{figure}

We explore the relevant region of the phase diagram of QCD 
for the Gaussian formfactor of our model. 
For  strong coupling constant $\eta=1$, 
a rich structure of color superconducting quark matter phases  
is found above the first order chiral transition (solid line on the left) 
as it is shown in Fig. \ref{phasediag}.  
For the quark matter sector we obtain that at low temperatures 
(up to $25 \div 35$  MeV depending on $\mu_B$) 
the mixed phase NQ-2SC is energeticaly  favored with increasing 
volume fraction $\chi$ of the 2SC phase as the temperature  increases 
(solid lines caracterize equal values of $\chi$ indicated over the lines). 
At intermediate temperatures a large region of pure 2SC phase is obtained. 
At higher temperatures 
a narrow window of g2SC occurs
before the second order phase transition to a deconfined phase takes place.
Nevertheless, for weak and intermediate coupling constants (e.g. for $\eta=0.75$ MeV) this structure of superconducting phases disappears and the normal quark matter phase 
is preferable. 

For higher values of the chemical potential, $\mu_B \ge 1500$ MeV, we expect the strange 
quark to appear and CFL phase to dominate 
(see  \cite{Phase:2005} for a recent three flavor phase diagram 
with a similar parametrization within an NJL model).

\section{Quark matter equation of state}

\begin{figure}[htb]
  \begin{center}
    \includegraphics[width=0.7\linewidth,angle = -90]{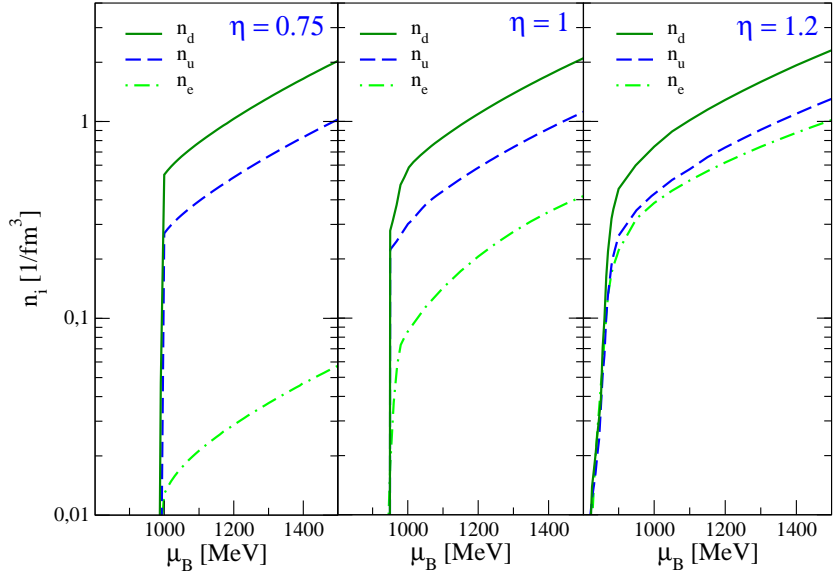}
     \vspace*{-0.5cm}
    \caption{Quark and electron number densities as a function of the baryon 
chemical potential at $T=0$. 
The three panels show results for different coupling constants of the 
diquark interaction, $\eta=0.75,~1.0,~1.2$~. }
     \vspace*{0.5cm}
    \label{numbers}
  \end{center}
\end{figure}

\begin{figure}[hbt]
  \begin{center}
    \includegraphics[width=0.7\linewidth,angle = -90]{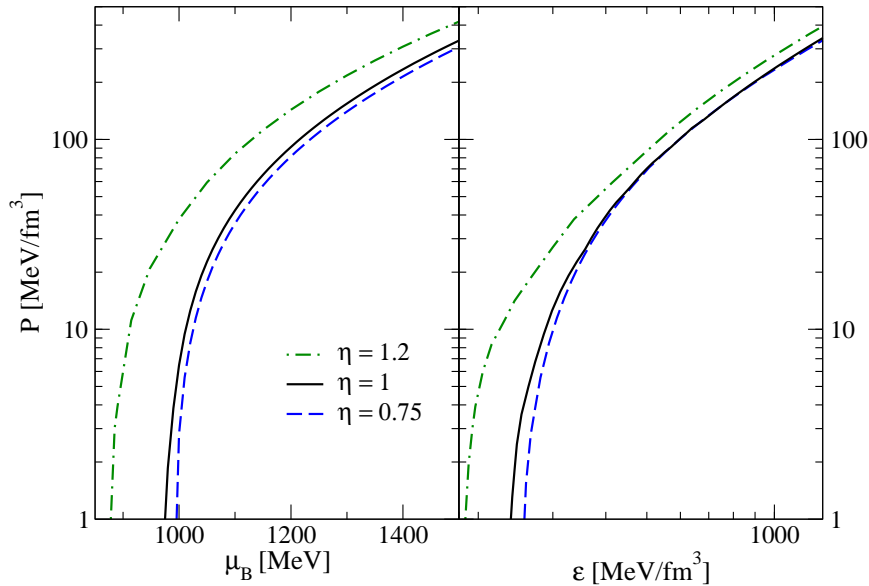}
     \vspace*{-0.5cm}
    \caption{ EoS for strongly interacting matter at zero temperature under
compact star constraints for three coupling parameters $\eta=0.75,~1.0,~1.2$~. 
Left panel: pressure vs. chemical potential; right panel: pressure vs. 
energy density.}
     \vspace*{0.5cm}
    \label{EoS}
  \end{center}
\end{figure}

The number density $n_j$ of the particle species $j$ in the mixed phase 
are given by
\begin{eqnarray}
n_j = \chi n_{j_{\Delta>0}}+ (1-\chi)n_{j_{\Delta=0}}~,
\end{eqnarray}
and shown as functions of $\mu_B$ at zero temperature in Fig. \ref{numbers}.
From these graphs we can see that  increasing $\eta$  the number of 
electrons in  the system increases by order of magnitudes and the onset 
of the phase transition from the vacuum is shifted to lower densities.
The energy density $\varepsilon$ in the mixed phase is given in a similar
form as
\begin{eqnarray}
\varepsilon = \chi \varepsilon_{\Delta>0}+ (1-\chi)\varepsilon_{\Delta=0}~.
\end{eqnarray}
We evaluate the quark matter EoS at $T=0$ within this nonlocal 
chiral model and show 
the results in Fig. \ref{EoS}.


\begin{figure}[hbt]
  \begin{center}
    \includegraphics[width=0.7\linewidth,angle = -90]{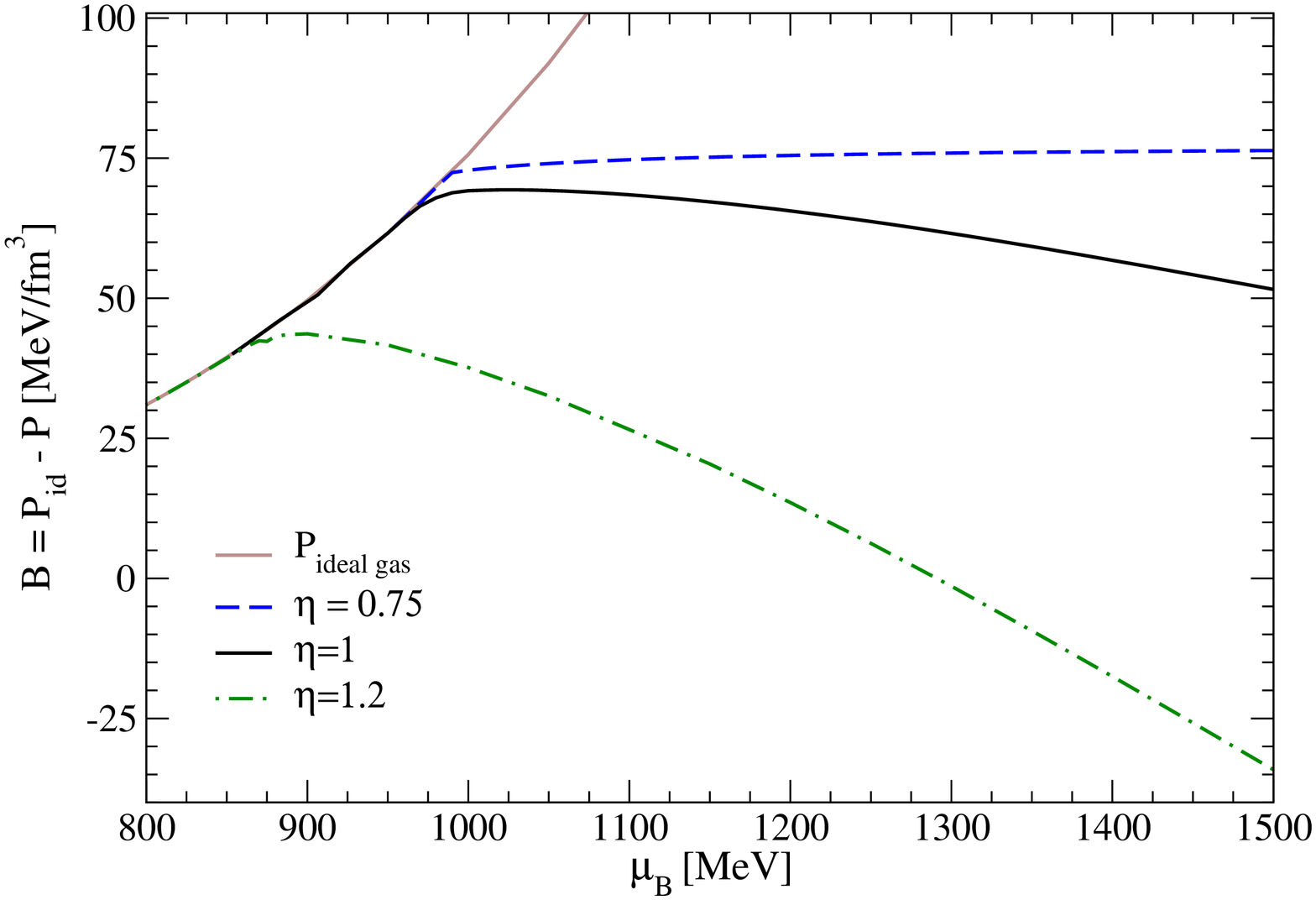}
     \vspace*{-0.5cm}
    \caption{ Bag pressure  - as defined in \cite{Grigorian:2003vi} - 
for different diquark coupling constants $\eta=0.75,~1.0,~1.2$
in dependence on the baryon chemical potential for Gaussian formfactor at $T=0$.  
}
     \vspace*{0.5cm}
    \label{bag}
  \end{center}
\end{figure}

As in \cite{Grigorian:2003vi}  we can  display the results for the pressure 
in the form of a bag model
\begin{equation}
\label{press}
  P = P_{id}(\mu_B) - B(\mu_B)~,
\end{equation}
where $P_{id}(\mu_B)$ is the ideal gas pressure of quarks and $B(\mu_B)$ a
{\it density dependent} bag pressure, see Fig. \ref{bag}. 
For $\eta=0.75$, $B\simeq 75$ MeV/fm$^3$ is nearly a constant function of 
the density; for $\eta=1$ we found a large pressure effect of the diquark 
condensate and therefore a lower and strongly density dependent $B$.  
Finally, for  $\eta=1.2$ the pressure of the condensate is so huge that the 
bag pressure  reaches negatives values.

\begin{table}[htb]
\begin{center}
\begin{tabular}{|c||c|c|c|}\hline
Formfactor   &$\Lambda$[GeV] &$G_1~\Lambda^2$ &$m$[MeV]\\ \hline
Gaussian   &$1.025 $ &$3.7805$&$2.41$\\
Lorentzian     &$0.8937$ &$2.436 $&$2.34$\\
NJL      &$0.9   $ &$1.944 $&$5.1 $\\ \hline
  \end{tabular}
\vspace{1cm}
  \caption{Parameter sets ($\Lambda$, $G_1~\Lambda^2$, $m$) of the nonlocal
chiral quark model for the different formfactors.
}
     \vspace*{0.5cm}
  \label{par}
\end{center}
\end{table}
\section{Conclusion}

Within our study of the equation of state and phase structure for 
two-flavor quark matter at low temperature under compact star constraints we
found that the occurence of a 2SC phase is sensitive to variations of 
both the formfactor of the nonlocal quark interaction and the strength of 
the diquark coupling constant. Our results suggest that for standard
values of the coupling $0.5<\eta<0.75$ either the 2SC phase does not
occur (Gaussian formfactor) or it exists only in a mixed phase with normal
quark matter with a volume fraction less than $20 - 40~\%$, occuring at high
baryon chemical potentials $\mu_B >1200$ MeV and most likely not relevant for
compact stars. 
We obtain that neither the g2SC phase occurs in compact stars; it appears in a narrow
window at higher temperatures and for strong coupling constants (e.g. $\eta=1$).

This result suggests that other pairing patterns like the  
CSL phase with relatively small gaps of the order of $10 - 100$
keV may be realized. This hypothesis is in accord with a recent 
description of modern cooling data using hybrid star models.

\section*{Acknowlegement}
We thank our colleagues for their  discussions and interest related to
our work, in particular J. Berdermann, M. Buballa, M. Huang, I. Shovkovy,
N.N. Scoccola and D.N. Voskresensky.  
The research of D.N.A. has been supported by 
Landesgraduiertenf\"orderung (MV-Germany), H.G. was supported by the 
Virtual Institute of the Helmholtz Association under grant No. VH-VI-041
and by the DAAD partnership programme between the Universities of 
Yerevan and Rostock. D.N.A. and D.B. acknowledge the hospitality of
Tandar Laboratory, CNEA, Buenos Aires during the final stages of this work and the
DAAD-ANTORCHAS programme for financial support of their visit under
grant No. D/04/27956.

\end{document}